# Multifractal random walk in copepod behavior

François G. Schmitt[a,*], Laurent Seuront[b]

[a] *Department of Fluid Mechanics, Vrije Universiteit Brussel, Pleinlaan 2, B-1050 Brussels, Belgium, francois@stro.vub.ac.be*

[b] *Ecosystem Complexity Research Group, Station Marine de Wimereux, CNRS UPRESA 8013 ELICO, Université des Sciences et Technologies de Lille, 28 avenue Foch, BP 80, F-62930 Wimereux, France*

[*] Corresponding author. Tel.: +32- 2 6292334; fax: +32- 2 6292880

**Abstract**. A 3D copepod trajectory is recorded in the laboratory, using 2 digital cameras. The copepod undergoes a very structured type of trajectory, with successive moves displaying intermittent amplitudes. We perform a statistical analysis of this 3D trajectory using statistical tools developed in the field of turbulence and anomalous diffusion. We show that the walk belongs to "multifractal random walks", characterized by a nonlinear moment scaling function for the distance versus time. To our knowledge, this is the first experimental study of multifractal anomalous diffusion. We then propose a new type of stochastic process reproducing these multifractal scaling properties. This can be directly used for stochastic numerical simulations, and is thus of important potential applications in the field of animal movement study, and more generally of anomalous diffusion studies.

# 1. Introduction

For most animals, movement behavior determines how individuals encounter features of their environments that vary in time and in space. Space-time variability in such ecological phenomena as foraging behavior, mating probabilities, population distribution, metapopulation dynamics, predator-prey or parasitoid-host interactions, or community composition, may therefore be mechanistically determined by how individual movement behavior is influenced by space-time heterogeneity in environmental features [1-3].

For zooplankton, competition, predation and aggregation occur across distances of centimeters to meters. However, the consequences of these interactions influence processes such as climate and fisheries productivity up to the global scale [4-6]. In this context, and considering the extremely intertwined properties of swimming and feeding processes in copepods ecology [7], to test mechanistic hypotheses that relate individual movements to higher-level ecological phenomena requires that individual swimming pathways be precisely characterized, both qualitatively and quantitatively.

In the following, we first propose a state-of the art in animal behavior studies, in both terrestrial and aquatic studies (section 2). We present our choice of copepod species and experimental procedure in section 3. Section 4 presents our statistical analysis of copepod swimming behavior. We experimentally show for the very first time that copepod displays a multifractal random walk, whose statistics are characterized by a nonlinear scaling exponent for the statistical moments of the increment of position. We also show that successive displacements display long-range power-law correlations, invalidating the usual uncorrelated random walk hypothesis. In section 5 we discuss the consequences of our findings and propose a new scaling stochastic process that reproduces the nonlinear scaling moment function for the displacements. In section 6 we give some final comments and directions of future developments.

## 2. Some previous animal behavior studies

### *2.1. Characterizing movement pathways*

Movement pathways have been characterized by a variety of measures, including path length (the total distance traveled, or gross displacement), move length (the distance traveled between consecutive points in time), move duration (time interval between consecutive spatial points), speed (move length divided by move duration), turning angle (the difference in direction between two successive moves), turning rate (turning angle divided by move duration), net displacement (the linear distance between starting and ending point), often used as a metric when making comparisons with diffusion or correlated random walk models [8-9], NGDR, i.e. Net to Gross Displacement Ratio [10], and fractal dimension [11-12]. These different metrics have been widely applied in both terrestrial and aquatic ecology. In the next subsections, we briefly exemplify the main ecological conclusions drawn from behavioral studies in both terrestrial and aquatic ecology.

### *2.2. Movement pathways in terrestrial ecology*

Characterizing the movement of organisms has a long lasting history [1], and a profusion of applications are found in entomological studies [2]. More specifically, Brownian and fractional Brownian motion models [12] has been proposed to mimic insect motions in a variety of environments. Dicke and Burrough [13] used fractal analysis to examine spider mite movements in the presence and absence of a dispersing pheromone. Following a different approach, Wiens and Milne [14] examined beetle movements in natural fractal landscapes. They found that observed beetle movements deviated from the modeled (fractional Brownian) ones. A follow-up study [9] found that beetle movements reflect a combination of ordinary (random) and anomalous diffusions. The latter may simply reflect intrinsic departures from randomness, or result from barrier avoidance and utilization of corridors in natural landscapes. Johnson et al. [15] discuss the interaction between animal movement characteristics and the patch-boundary features in a 'microlandscape'. They argue that such interactions have important spatial consequences on gene flow, population dynamics and other ecological processes in the community [16]. In a comparison of path tortuosity in three species or grasshopper, With [17] found that the path fractal dimension of the largest species was smaller than those of the two smaller ones. She suggested that this reflects the fact that smaller species interact with the habitat at a finer scale of resolution than do larger species. A subsequent study [18] showed differences in the ways that gomphocerine grasshopper nymphs and adults interacted with the microlandscape.

Following the development of observational methods such as radio-telemetry [19], similar studies have been conducted to characterize the search paths of small [19-22] and large [23-25] mammals. These behavioral studies then provided salient informations to identify specific food search-strategies from local to global scales. For instance, finch flocks tend to move forward rather than turn to the side, and almost never turned backward [20], red foxes moves during the night and remain relatively stationary during the day [19], and wolf females movement paths show significant changes throughout the year, depending on the state of their life cycle (normal, breeding and wandering) [23]. Similar studies are also essential to discriminate between (i) acquired behavior, perceptible from the spatial memory of vervet monkeys who appear to behave as though they can "look ahead" at least three steps and thus might use a permutational heuristic to solve spatial foraging problems [21], and (ii) inborn behavior related to straighter movements of caribou during their annual migrations [25].

## 2.3. Movement pathways in aquatic ecology

Despite a fast growing number of studies, analyses of movement behavior of aquatic organisms are less common in aquatic than of terrestrial organisms, primarily because of the difficulty in obtaining accurate records of the displacement of swimming organisms, which unlike terrestrial organisms, take place through a volume and therefore require systems capable of recording three-dimensional (3D) data. Taking advantages of the recent improvements of radio-telemetry technology, approaches devoted to study the behavior of large animals such as sharks [26-27], tuna [28], sea turtles [29], seals [30] or marine birds [31-33] showed an important development in the last ten years. It is then now feasible to get records down to several hundreds of meters, over wide geographic areas and for several months.

On the other hand, behavioral studies of smaller animals such as tiny fishes, crabs and plankton organisms required the use of sophisticated *in situ* [34-35] and laboratory [36] video system. Examples come from the wide spectrum of swimming behaviors related to the species [37], the age [38-41], the prey density [42-44], the presence of a predator or a conspecific [41, 45-46], the sex of individuals [39,47-48], the information imparted into the surrounding water by a swimming animal [49-50], including both chemical [51-52] and hydromechanical [47, 53-56] stimuli. In particular, as widely demonstrated for terrestrial organisms, zooplankton organisms are now recognized to present behavioral adaptation to resource patchiness: they increase [39] or decrease [42-43, 57] their swimming speed with increasing phytoplankton densities, i.e. a species-specific adaptation to increase the encounter rate with their preys, and increase the tortuosity of their swimming paths in food patches [57-60].

However, one needs to note here that the development of such video recording apparatus is a non-trivial problem, especially in plankton ecology, and has resulted in many investigations of zooplankton behavior recording only two-dimensional (2D) swimming pathways. Moreover, even using video systems capable of recording 3D data, there are still problems of scale resulting from the small size of planktonic organisms. Gathering 3D coordinates for zooplankton involves a trade-off between resolution and extent, typically presenting researchers with two alternatives in the collection of data: (i) high spatial resolution, but for short duration, or (ii) longer time series, but at low spatial resolution. To our knowledge, only two studies have permitted the collection of 3D swimming data at both high spatial resolution and for long periods [38, 43]. We then present in the following an alternative, original and simple procedure to acquire reliable three-dimensional data of the swimming behavior of zooplankton organisms.

## 3. Recording three dimensional behavior of swimming organisms

### *3.1. Living material collection and acclimation*

The copepods are the largest and most diversified group of crustaceans, and are the most numerous metazoans (i.e. all multi-celled organisms) in the aquatic communities. At present they include over 14.000 species, 2.300 genera and 210 families, a surely underestimated number. Their habitat ranges from fresh water to hyper saline conditions, from subterranean caves to water collected in bromeliad leaves or leaf litter on the ground, from streams, rivers, and lakes to the sediment layer in the open ocean, from the highest mountains to the deepest ocean trenches, and from the cold polar ice-water interface to the hot active hydrothermal vents. Copepods may be free-living, symbiotic, or internal or external parasites on almost every phylum of animals in water. The usual length of adults is 1-2 mm, but adults of some species may be as short as 0.2 mm and others may be as long as 10 mm. They are considered the most plentiful multicultural group on the earth, outnumbering even the insects, which include more species, but fewer individuals. Particularly, the copepods are the dominant forms of the marine plankton, constitute the secondary producers in the marine environments and then a fundamental step in the oceanic food chain, linking microscopic algal cells to juvenile fishes and whales. Copepods also have the potential to act as control mechanisms for malaria by consuming mosquito larvae, and contrariwise are intermediate hosts of many human and animal parasites.

In this preliminary study, we focused on the carangid copepod *Temora longicornis* (Figure 1), a very abundant species in worldwide coastal waters, which is also of great ecological significance in many areas. For instance, it represents 35 to 70% of the total copepod population in the southern Bight of the North Sea [61], and in Long Island Sound, USA, *T. longicornis* has been shown to be able to remove up to 49% of the daily primary production [62]. Feeding and swimming being two intertwined processes in copepods ecology, the precise characterization of the swimming behavior represent a salient issue in marine ecology.

Individuals of the copepod *Temora longicornis* were collected with a WP2 net (200 mesh size) in the offshore waters of the Eastern English Channel. Specimens were diluted in buckets and transported to the laboratory. The acclimation of copepods consisted of being held in 20 l beakers filled with 0.45 µm filtered seawater to which was added a suspension of the diatom (a phytoplankton algae) *Skeleton postpartum* to a final concentration of $10^8$ cells.$l^{-1}$. Prior to the filming experiment, adult females were sorted by pipette, acclimated for 24 h at 18°C and fed on a mixture of the green algae *Nannochloropsis occulata* (3 µm) and the flagellate *Oxyrrhis marina* (13 µm). The larger heterotrophic flagellate *Oxyrrhis marina* was present as an additional food source. In this preliminary experiment, an adult female (1.1 mm) was sorted by pipette and left in the experimental filming set-up to acclimatize for about 15 mn prior to filming. A realistic concentration of the mixture of *Nannochloropsis occulata* and *Oxyrrhis marina* was tested: *Nannochloropsis occulata* and *Oxyrrhis marina* at $10^8$ and $10^6$ cells.$l^{-1}$, respectively.

## *3.2. Experimental design*

The experimental set-up (Figure 2) was designed to track the three-dimensional displacement of a copepod in a cubic glass container (inner side 15 cm, effective volume 3.375 l). Illumination came from two lamps (diffuse cold light 75W) located above and below the container to ensure homogeneity of the light source and thus to avoid phototropism. Reflections were minimized as much as possible and two lateral sides of the container were covered with black dull plastic film to increase the contrast between the copepod and the container background.

The three-dimensional trajectory of the copepod was recorded using two orthogonally focused and synchronized CCD cameras (HITACHI KP M1; 875*560 pixels; focal distance 17:53 mm), facing the two observations frames of the experimental container. The cameras deliver black and white frames at a rate of 12.5 frames per second, and rulers provide the relation between pixels and millimeters. An encoder RGB-PAL [Enc110 (For-A)] codes PAL-type red and green frames from the two instantaneous synchronized monochrome images. Each orthogonal view then gets an identity and may be added to another at the same time $t$ to form one single color PAL-frame. In particular, this frames superposition reduces hardware cost and gives perfect synchronization. The color images were digitized (720*576 pixels), compressed and stored in real time using a special acquisition card and an appropriate software (PVR-Digital Processing Systems) on a PC.

Finally, the three components of the copepod trajectory were extracted using frame analysis. A pre-recorded background was removed from each frame, leading to a decomposition of the initial colored frames into gray-level frames giving two binary frames by thresholding. After a morphological opening to filter parasitic points, each binary frame is labeled to get the coordinates (in the image system) and the objects numbered on it. This frame processing was carried out on the acquisition PC (Pentium Pro 200 MHz) at a rate of 1.5 seconds per frame. An identification software subsequently eliminates spurious objects (i.e. ludicrous discontinuities in the trajectory). Another software takes into account spatial parallax and diffraction phenomena and gives the three coordinates of the copepod as a function of time in the container coordinates system. While the frame-disk is able to store about 40 minutes of a series at 12.5 images per second, we only recorded the swimming behavior of the *Temora longicornis* individual for 7 minutes, after which the three-dimensional trajectory was reviewed and valid segments were identified for analysis. Valid segments consisted of pathways in which the animals were swimming freely, at least two body lengths away from any of the chamber's walls or the surface. In further analysis we use the longer (i.e. 2 minutes 42 seconds) valid segment available from our lab experiment.

### *3.3. Behavioral observations*

The three-dimensional record of the swimming behavior of *Temora longicornis* is shown in Figure 3. *T. longicornis* moved actively in a highly variable and irregular pathway, showing an alternance between periods of relative straight swimming and periods of erratic motions including strong jumps in all three dimensions, ensuring the three-dimensional nature of the recorded pathway. Both the qualitative and quantitative nature of these fluctuations will be investigated in the next section using an original statistical procedure.

## 4. Statistical analysis of the copepod displacements

Strictly speaking, planktonic organisms are passively advected by the surrounding water masses, or present swimming abilities that are neglectible when compared to the amplitude of turbulent motions. In practice copepod displacements result of their own movements, that must be added to the advective motions generated by oceanic turbulence. Here we specifically study copepod movements without turbulence, in order to be able to directly characterize its specific movement ability and to establish a biological reference framework, before considering more complex situations where the physical effects of turbulence affect biological properties. As we shall see, contrary to what could be believed, not only copepods are able to swim very fast relative to their size and to the turbulence levels they experience in the field, but also to produce a very structured type of movement.

After comparing copepod swimming velocity and turbulent characteristic velocity, we analyze here the statistics of copepod displacements using an original approach coming from the field of anomalous diffusion. We first characterize the multifractal properties of three-dimensional copepod displacements, then consider their two-dimensional projections. We further analyze the statistics of successive displacements: their amplitudes and angles.

### *4.1 Copepod swimming speed and turbulence*

We present here a brief comparison between the swimming ability of the copepod *Temora longicornis* and the turbulent velocity relevant at their characteristic scale. The root-mean-square turbulent velocity $w$ (referred as rms turbulent velocity hereafter) at a scale $d$ writes $w = C\varepsilon^{1/3}d^{1/3}$ where $C$ is a constant and $\varepsilon$ (m$^2$.s$^{-3}$) is the small-scale turbulent dissipation rate belonging to the inertial subrange. At the scale of a plankton organism $l$(m), the most relevant turbulent velocity can then be expressed as [63]:

$$w = 1.37\varepsilon^{1/3}l^{1/3} \qquad (1)$$

Here, considering the characteristic scale $l$ =1.1 mm (i.e. the body length of the copepod), we estimate the rms turbulent velocity $w$ for different values of the dissipation rate $\varepsilon$ ranging between $10^{-10}$ and $10^{-4}$ m$^2$s$^{-3}$. Values bounded between $10^{-10}$ and $10^{-6}$ m$^2$s$^{-3}$, $10^{-7}$ and $10^{-6}$ m$^2$s$^{-3}$ and $10^{-6}$ and $10^{-4}$ m$^2$s$^{-3}$ are characteristic of the open ocean, the continental shelf and highly dissipative coastal waters, respectively. It can be seen from Figure 4 that the observed swimming velocity $v$ of *Temora longicornis*, i.e. v=1.3 mms$^{-1}$ is larger that the rms turbulent velocity observed in most marine environments. Indeed, the rms velocity overcomes the swimming velocity of the copepod *T. longicornis* only for very high values of the dissipation rates $\varepsilon$ ($\varepsilon \geq 10^{-5}$ m$^2$s$^{-3}$) characterizing extremely turbulent environments such as breaking waves or narrow tidal channels. However, the maximum swimming velocity, i.e. v=29.0 mms$^{-1}$, is significantly higher (an order of magnitude) than the rms turbulent velocity observed in the highest turbulent marine areas. These very simple arguments then demonstrate that the concept of "plankton" is all relative and should be used very carefully even when applied to swimming animals as tiny as copepods.

### *4.2 Multifractal study of copepod displacements*

We note here $\mathbf{X}(t)$ the three-dimensional position of the copepod at time t. We consider the norm of its displacements $\Delta \mathbf{X}_\tau = \mathbf{X}(t + \tau) - \mathbf{X}(t)$: by hypothesis, the moments of order q>0 of the displacements depend only on the time increment $\tau$. In case of scaling, we introduce the moment function exponent $\zeta(q)$ defined as [64-70]:

$$<\|\Delta \mathbf{X}_\tau\|^q> \approx \tau^{\zeta(q)} \qquad (2)$$

This exponent function is very useful to characterize the statistics of the random walk. For Brownian motion, it is well-known that $\zeta(q)=q/2$. Classically, only the moment of order 2 is estimated, and whenever $\zeta(2)=1$, the process correspond to a normal diffusion, whereas the rich field of anomalous diffusion corresponds to dispersive processes with $\zeta(2) \neq 1$ (see reviews on anomalous diffusion in [71]). The idea behind this characterization using only one moment was implicitly to assume that if $\zeta(2)=1$, one has for all qs $\zeta(q)=q/2$, so that the process has the same diffusive properties as Brownian motion. Of course, this is not necessarily the case, and one can have $\zeta(2)=1$ (so that the diffusion is apparently normal, see an example in [72]) whereas for other moments $\zeta(q) \neq q/2$. It is thus better to characterize the process with the whole function $\zeta(q)$ instead of a single exponent, as done in Refs. [64-70]. In this framework, it would be more coherent to denote "anomalous diffusion" as a diffusion with a function $\zeta(q) \neq q/2$.

When the function $\zeta(q)$ is nonlinear, we refer the resulting diffusion as being "multifractal", by analogy with multifractal characterization of correlations and intermittency in turbulence (see [73-74] for recent reviews). The terms "multifractal random walk" were already proposed in [70]. Before this, a diffusion characterized by a nonlinear $\zeta(q)$ function has received different names: generalized diffusion [64], multidiffusion [66], multifractionnal kinetics [67], or strong diffusion [68-69].

We first analyzed the three-dimensional trajectory of the copepod movements shown in Figure 3. The scaling of several moments is shown in Figure 5. It can be seen that the scaling is very good for a range of scales of 2 decades, for times between 0.3 and 30 s, corresponding to distances between 0.4 and 4 cm (the distance goes as the square-root of time, so that a time ratio of 100 corresponds to a distance ratio of 10). For smaller distances the departure from scaling could be due to the reaction time of the copepod. In order to precisely detect a departure from normal diffusion, this scaling behavior is displayed in Figure 6 with a compensation by the normal diffusion scaling exponent: this shows the anomalous diffusion obtained for the range of scales presented above. It can be seen that $\zeta(2)$ is slightly smaller than 1, with a very small correction; the correction is larger for q>3. In the following, scaling exponents are estimated as a least-square power-law slope. Figure 7 shows the resulting $\zeta(q)$ function, which is clearly nonlinear. The straight line of equation q/2 is shown for comparison. For moments larger than 2, the departure from this straight line is very clear. The function obtained is nonlinear and convex, as opposed to the numerical studies presented in [68] where the same function is bilinear and concave. This characterizes different types of anomalous diffusion.

We also compare the functions obtained in three dimensions with the scaling exponents estimated from two-dimensional projections in the xy, xz and yz plans. This can be useful for comparison with other databases where only 2 coordinates are recorded. For example a numerical study in two dimensions has been performed in [72], where the copepod was assumed to diffuse in a multifractal phytoplankton field. Figure 8 shows that the $\zeta(q)$ functions estimated from two-dimensional data have the same shape as the original three-dimensional one, with some statistical variations which are expected to be due to the finite size of the data base.

### *4.3 Analysis of successive displacements: amplitudes and angles*

Here we perform some complimentary analysis, helping to provide some general understanding of the structure of the copepod random walk. We consider the "moves" as the difference in position of the copepod for time increments of 0.08 s (the resolution time step). We consider the structure of these elementary moves: the correlation of their amplitude, and the angle between successive moves.

We display the successive move amplitudes in Figure 9: it can be seen that the moves are very intermittently distributed, with some large ones separated by very small ones. This time series has a specific structure, recalling the fields resulting from direct multiplicative cascades (see [73]). Its correlation function, represented in Figure 10, shows long-range power-law correlation. This behavior is in agreement with the nonlinearity of the ζ(q) function. We also considered the angle between successive moves. No correlation could be detected between the angle and the amplitude of moves. The angles themselves are not equally probable: Figure 11 shows their histogram, indicating as expected, that angles larger than 90 deg are less probable: the copepod goes more likely forward than backward. Nevertheless, it goes sometimes backward, and the histogram shows that a local maximum is obtained for angles of about 20 deg: the copepod goes more likely ahead of him with a small deviation in angle. These are interesting characterization that could be taken into account when simulating copepod behavior in e.g. multi-agent type of simulations [75].

## 5. Discussion and proposal of a new scaling stochastic process

We have obtained above experimental evidence of anomalous diffusion for three-dimensional data, with multifractal statistics. For theoretical as well as numerical developments of these results, the question then arises of how to formalize or simulate stochastic processes possessing such multifractal random walk properties. We discuss here briefly this question. We first provide a general framework possessing the required multiscaling statistics, then we discuss of possible stochastic kernels available. We also discuss other published results in the same direction.

### *5.1 Stochastic processes for multifractal random walks*

Let us consider first the one-dimensional case. Possible three-dimensional generalizations will be discussed elsewhere. Let us consider a stochastic kernel $f(t,s)$ possessing the following properties:

$$f(at, as) \stackrel{d}{=} g(a)f(t,s) \quad a > 0 \tag{3}$$

where '$\stackrel{d}{=}$' means equality in distribution, and g(t) is a non-stationary random process characterized by its moments q>0:

$$<|g(t)|^q> = t^{-\psi(q)} \quad t > 0 \tag{4}$$

where $-\psi(q)\ln t$ is its second Laplace characteristic function, assumed to converge (at least for some moments q>0). We take here (3) and (4) as granted, and discuss of how to built such process in the next sub-section.

We then define the random walk as follows:

$$X(t) - X(0) = \int_0^t f(t,s)dB(s) \tag{5}$$

where B(t) is a Brownien motion. When f=1 (non-random) this recovers the classical random walk for X(t). In the general case, this walk possesses the following scaling:

$$\begin{aligned}
X(at) - X(0) &= \int_0^{at} f(at,s)dB(s) \\
&= \int_0^t f(at,as')dB(as') \quad (s' = s/a) \\
&\stackrel{d}{=} \int_0^t g(a)f(t,s')a^{1/2}dB(s') \\
&\stackrel{d}{=} g(a)a^{1/2}(X(t) - X(0))
\end{aligned} \tag{6}$$

When taking t=1 fixed and a variable, and taking the moments of order q>0 of each side, one obtains:

$$<|\Delta X(a)|^q> = C_q a^{q/2} <|g(a)|^q> \approx a^{\zeta(q)} \tag{7}$$

where $C_q$ is a constant and:

$$\zeta(q) = \frac{q}{2} - \Psi(q) \tag{8}$$

We then discuss of how to obtain conditions (3) and (4) with a nonlinear function $\Psi(q)$.

## 5.2 Scaling stochastic kernels

Let us consider x(t) a process with stationary independent increments such as x(0)=0. Then for τ>0 it is well known that $y_\tau(t) = x(t+\tau) - x(t)$ is an infinitely divisible stochastic process (see e.g. [76]). Introducing—when defined—the second Laplace characteristic function

$$\Psi_{y_\tau}(q) = -\log < e^{qy_\tau(t)} > \tag{9}$$

one has:

$$\Psi_{y_\tau}(q) = \tau \Psi_{y_1}(q) \tag{10}$$

This may be rewritten as:

$$< e^{qx(t)} > = e^{-t\psi(q)} \tag{11}$$

where $\Psi(q) = \Psi_{y_1}(q)$. Let us introduce $g(t) = e^{x(\log t)}$. The moments of this process thus verify

$$< g(t)^q > = < e^{qx(\log t)} > = t^{-\Psi(q)} \tag{12}$$

so that one has, for a>0:

$$g(at) \stackrel{d}{=} g(a)g(t) \tag{13}$$

Now, taking $f(t,s) = g(|t-s|)$ for $t \neq s$, one obtains a stochastic kernel f(t,s) verifying relations (3) and (4), where $\Psi(q)$, as the second Laplace characteristic function of an infinitely divisible random variable, is a nonlinear and convex function.

This shows that any infinitely divisible probability distribution can be used as a basis of the stochastic kernel f(t,s) used to introduce a multifractal random walk. Let us mention among possible kernels the Poisson process, or the Levy stable family, among which is the Gaussian process. In the latter case, one has $\Psi(q) = Cq^2$ so that a possible form of $\zeta(q)$ would be:

$$\zeta(q) = \frac{q}{2} - Cq^2 \tag{14}$$

In Figure 7 a fit of the experimental curve of the form $\varsigma(q) = Aq - Cq^2$ was performed, with A=0.525 and C=0.028, shown for illustration. This provides a very good fit until a moment of order 6, which is already high for such a small data set. This is shown only for illustration purpose: it could well be that the multifractal diffusion is driven by another type of stochastic kernel (such as log-stable): there are too few data to investigate this point here.

### *5.3 Comparison with other proposals*

There have been several interesting proposals in the literature to simulate multifractal fields having the same scaling properties as turbulent velocity: Juneja et al. [77] have proposed a discrete (in scale) decomposition, and Biferale et al [78] a discrete (in scale) but causal process, that could be adapted to simulate a discrete multifractal random walk. These models possess discrete dilatation invariance, while Bacry, Delour and Muzy [70] have introduced a multifractal random walk model possessing continuous dilatation invariance properties. Nevertheless, they could only propose explicitly a discrete version of this model: they showed that the model is not ill-defined in the continuous limit, but could not provide the corresponding stochastic process. Furthermore, their approach corresponds only to the Gaussian case, while our proposal is more general, since it involves the whole infinitely-divisible family. Apart from these differences, the approach of [70] seems to be in the same direction as our proposal: a Gaussian stochastic integral with an independent stochastic kernel.

Our multifractal random walk model was presented here as an illustration of a process able to reproduce the main statistical properties shown by the data; a more detailed study of its properties will be discussed elsewhere.

## 6. Concluding remarks

Let us summarize here the main results of this paper, and breifly discuss their potential implications. We have presented a three-dimensional data base of a copepod random walk in a non-turbulent environment, and shown its multifractal characteristics. The present data base concerned only one realization, with 2 minutes and 42 seconds of data recorded at 12 Hz. In particular, the results presented here have several potential implications on our understanding of zooplankton behavior and trophodynamics. For instance, is the observed behavior an inborn property, or is a specific response induced by the environment?

In that way, one need to note that the behavior observed here could be related to the heterogeneous nature of phytoplankton distribution is in the experimental container. After being introduced in the cubic container, phytoplankton cells have been stirred by turbulent mixing. This turbulent mixing may well have produced a multifractal distribution of phytoplankton (see e.g. [79] for a multifractal study of the heterogeneous distribution of phytoplankton in the ocean). The container was then kept at rest for 15 minutes before the introduction of the copepod, so that the heterogeneous distribution is homogenized by molecular diffusion. A characteristic time scale of 15 minutes corresponds to scales homogenized by molecular diffusion up to 3 cm, which is close to the largest scale of our anomalous scaling range. The observed multifractal random walk then occurred over a range of scales where phytoplankton distribution is thought to be homogenized, and could not be related to a heterogeneous distribution of phytoplankton cells. This result then rather suggests an inborn rather than an induced behavior.

On the other hand, the potential identification of different modes in the swimming behavior of *Temora longicornis* (cf. Figure 8) in the vertical and the horizontal dimensions may reflect (i) the effect of gravity which affects most copepod species and *T. longicornis* in particular [36], and/or (ii) an adaptive reminiscence of diel vertical migration as a predator avoidance strategy [80], and/or (iii) a feeding switch between two different kind of food sources as the non-motile alga *Nannochloropsis occulata* and the motile flagellate *Oxyrrhis marina* used in the present study. Differential swimming behaviors in the horizontal and the vertical plans may also be suggested as a potential basis to investigate the predation risk associated with differential swimming behavior related to mating, feeding or predator avoidance strategies.

Whatever that may be, the full understanding of the very structured nature of the copepod's movements need further experimental work. In particular, a comparison of the behavior of different species of copepods will be done in the framework introduced here, as well as a study of the influence of the feeding conditions (i.e. quality and quantity of food) on the statistics of the walk. Moreover, let us note that a multifractal characterization of copepod spatial distribution has already been demonstrated in another context in turbulent oceanic conditions [81]. Further studies could then in particular be able to relate the statistics of copepod diffusion in real conditions with those of the spatial density of copepods (i.e. to relate lagrangian diffusion with the resulting spatial statistics) as previously suggested in terrestrial ecology [8].

These results were, to our knowledge, the first experimental evidence of multifractal anomalous diffusion: all the previous studies on this topic seem to be either theoretical or numerical simulations of different deterministic or stochastic diffusion models.

These results characterize copepod diffusion in the absence of turbulence. It provides preliminary information that can be used for numerical simulations as well as for theoretical models. Multi-agent copepod diffusion could also be analyzed in these lines, to detect if it provides a realistic enough simulation displaying multifractal statistics.

We have finally proposed a new and simple stochastic model, able to display such multifractal anomalous diffusion. This process is continuous in time, and in dilatation invariance (in scale), and general enough to include all infinitely divisible models. This model will be studied more thoroughly elsewhere.

## Acknowledgements

The authors are indebted to Dr. L. Falk and Dr. H. Vivier for letting us to use their lab facilities (Laboratory of Chemical Engineering Sciences, ENSIC, UPR CNRS 6811, Nancy, France); to Dr. P. Pitiot for his priceless help in the data acquisition and preprocessing; and to Ms. V. Denis for technical assistance. Thanks are also extended to the captain and the crew of the NO 'Sepia II' for their help during the sampling experiments, and to Prof. Y. Lagadeuc for initiating the collaboration with the Laboratory of Chemical Engineering Sciences.

# Figure Captions

Fig. 1: Dorsal (A) and side (B) view of the copepod *Temora longicornis*. Antennules (An) and antennae (A) hold chemo- and mechanoreceptors used to scan the surrounding environment. The cephalothorax (Ce) holds a rudimentary eye, the ocellus, and the mouth appendages (Bu) (mandibles, maxillules and maxillae) responsible for the capture and handling process of food particles. The thorax hold biramous swimming legs, the peraeopod (Pe). The abdomen (Ab) and the furca (Fu) do not hold any appendages, and the former holds the genital apparatus. The scale bars represent 0.5 mm.

Fig. 2: Schematic representation of the data acquisition system. The copepod is kept in a cubic glass container, and its swimming pathway is recorded using two synchronized and orthogonally focused cameras. After encoding of the two instantaneous synchronized images, the three components of the copepod trajectory were extracted using frame analysis.

Fig. 3: Three-dimensional swimming pathway of the copepod *Temora longicornis*. *T. longicornis* moved actively in a highly variable and irregular pathway, showing an alternance between periods of relative straight swimming and periods of erratic motions including strong jumps in all three dimensions. The black and gray arrows indicate the beginning and the end of the swimming path, respectively.

Fig. 4: Log-log plots of the root-mean-square turbulent velocity $w$ (mm.s$^{-1}$) estimated at the copepod characteristic scale (1.1 mm) as a function of the dissipation rate of turbulent kinetic energy $\varepsilon$ (m$^2$.s$^{-3}$). The dotted and dashed lines correspond to the mean and maximum swimming speed of the copepod *Temora longicornis*, v=1.3 mms$^{-1}$ and v=29.0 mms$^{-1}$, respectively, as observed in our lab experiment.

Fig. 5: The moments $<\|\Delta \mathbf{X}_\tau\|^q>$ vs. $\tau$ for q=1,2,3. A very nice scaling behavior is visible on 2 decades, for temporal increments between 0.3 s and 30 s.

Fig. 6: The compensated moments $<\|\Delta \mathbf{X}_\tau\|^q>/\tau^{q/2}$ vs. $\tau$ for q=1,2,3. A scaling of the residual indicates anomalous scaling diffusion, which is confirmed by the visible trends for the same temporal increments as Fig. 4.

Fig. 7: The function ζ(q) experimentally determined from the data (dark squares), compared to a fit corresponding to normal diffusion (q/2, dotted line) and a nonlinear multifractal fit (continuous line).

Fig. 8: The function ζ(q) experimentally determined from 3D data, compared to the functions estimated from 2D projections in the xy, xz,yz plances respectively.

Fig. 9: The series of amplitudes of successive moves, showing an intermittent structure.

Fig. 10: Autocorrelation function of the series of successive moves; confirming the long-range correlations of the displacements.

Fig. 11: Histogramme of the angle between successive displacements. The copepod goes sometimes backward. There is a local maximum at a non-zero angle of about 20 deg.

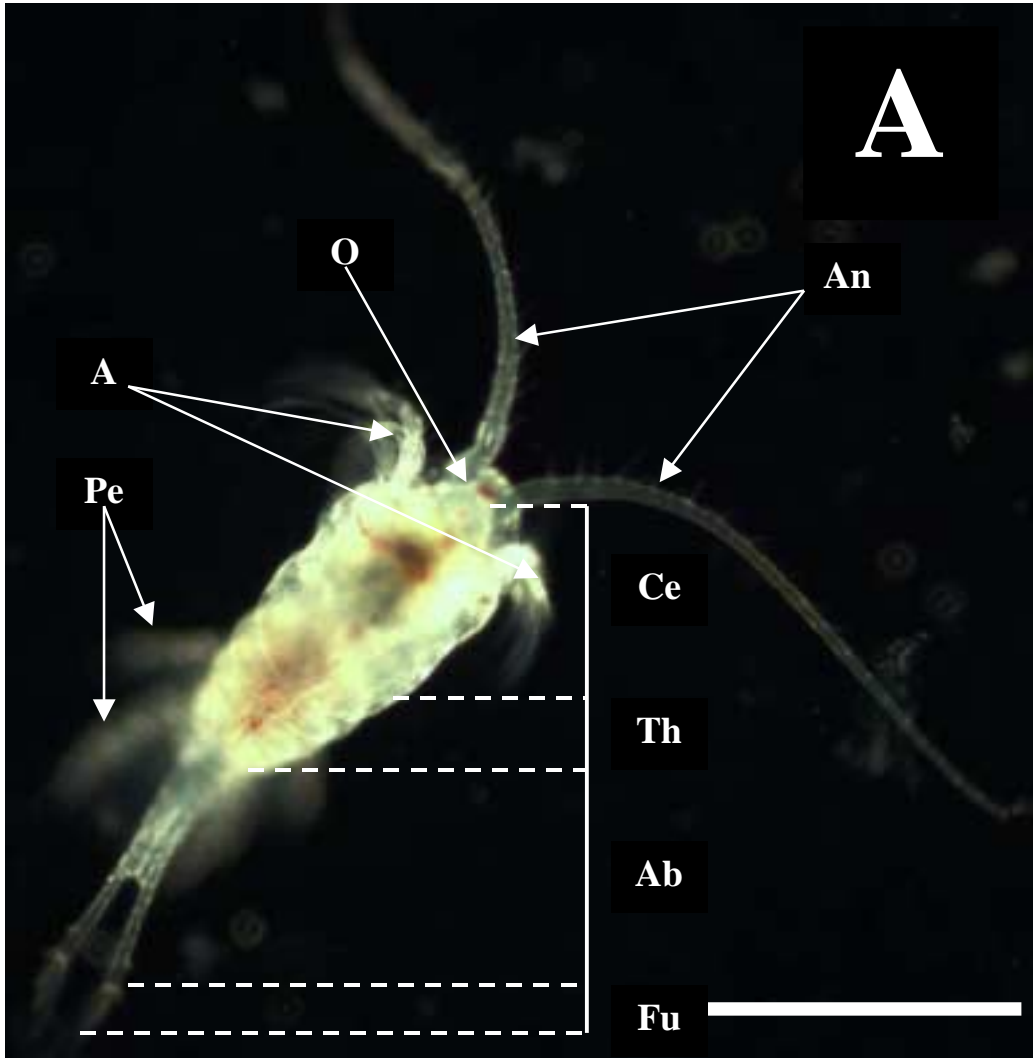
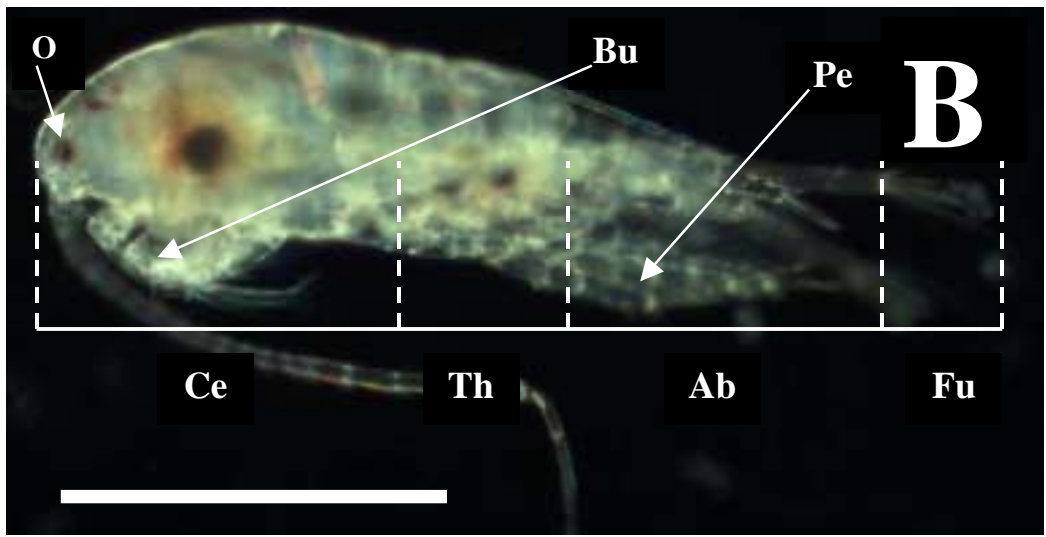

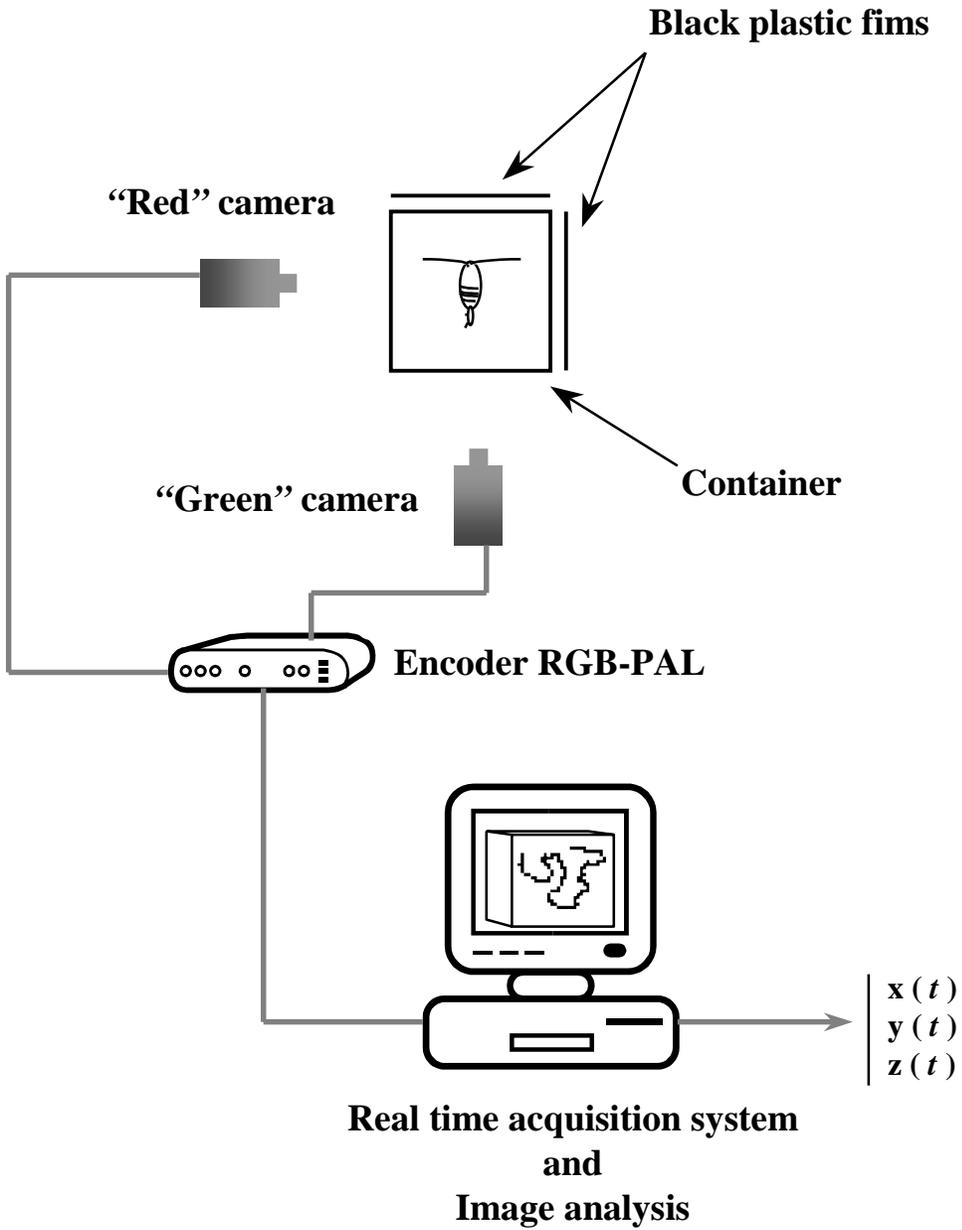

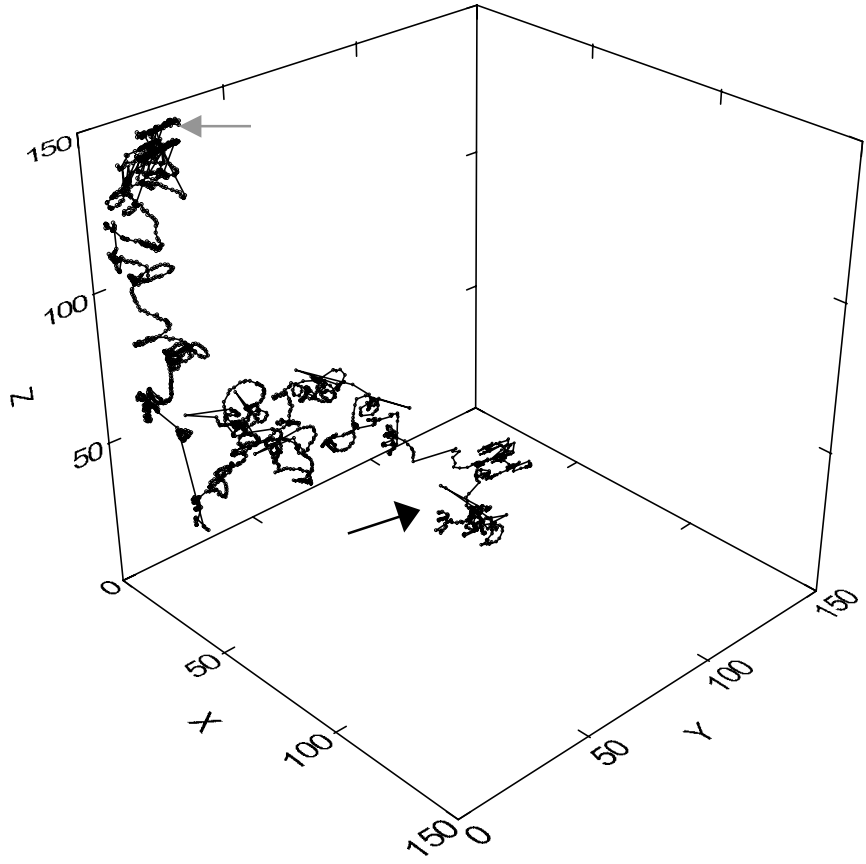

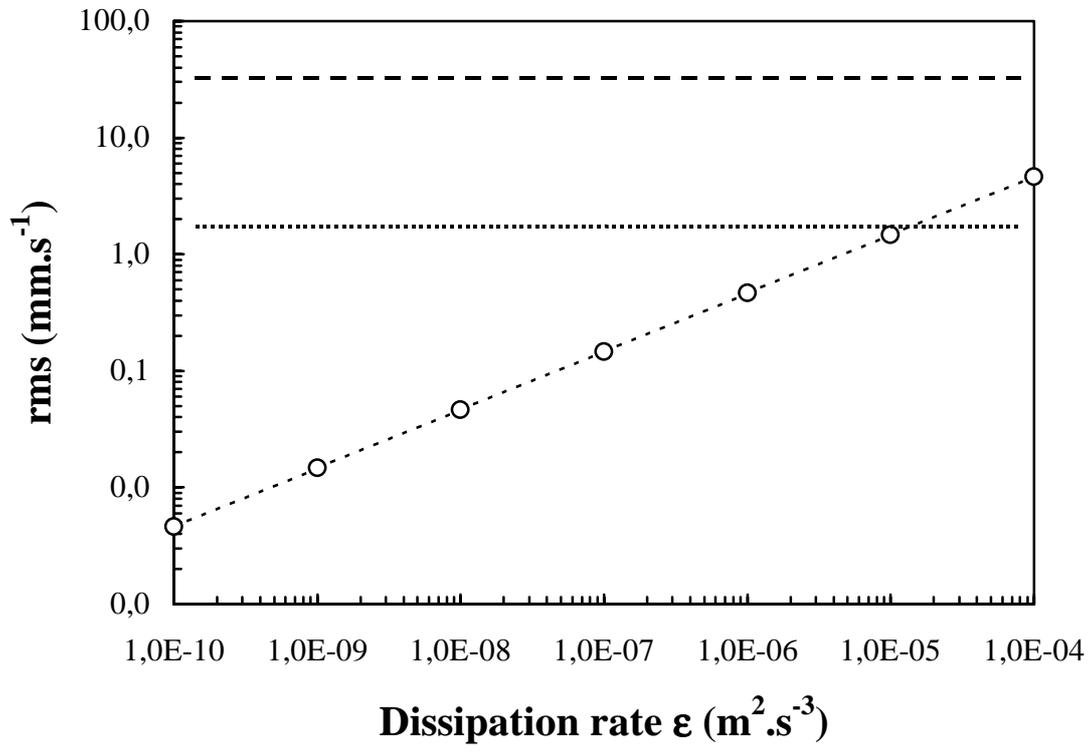

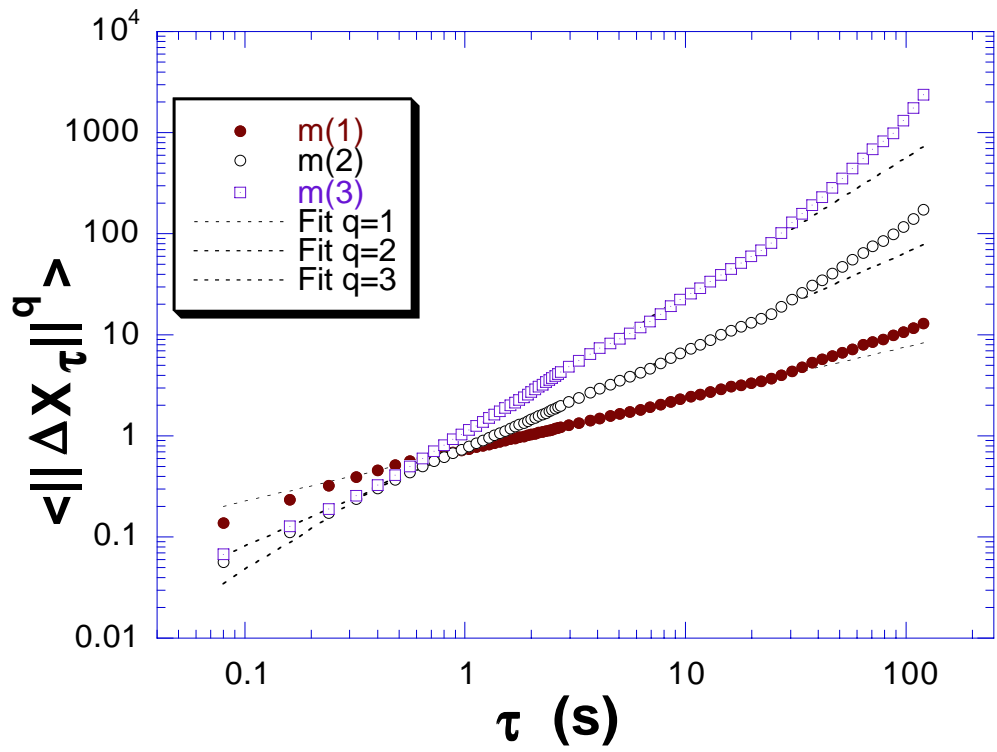

Fig. 5

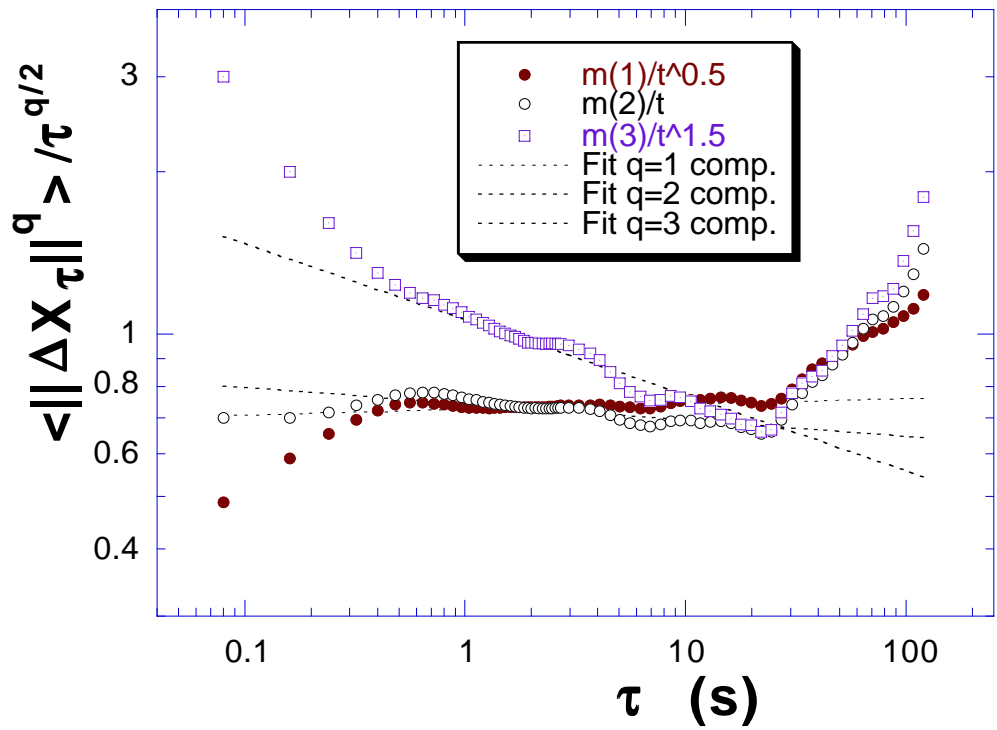

Fig. 6

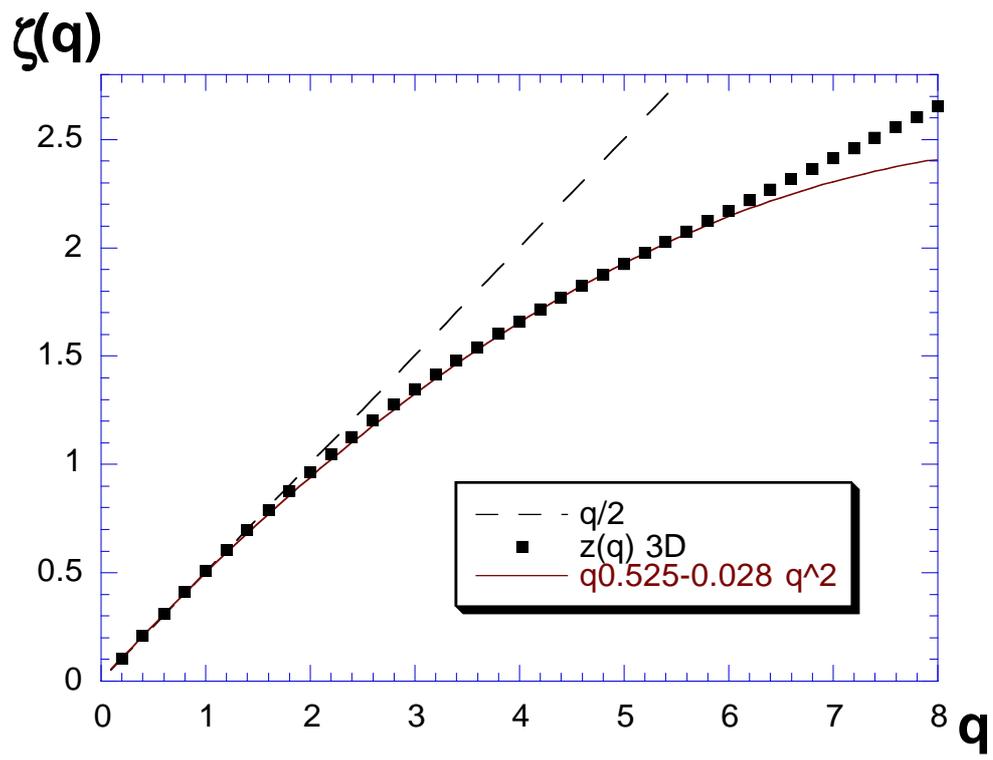

Fig. 7

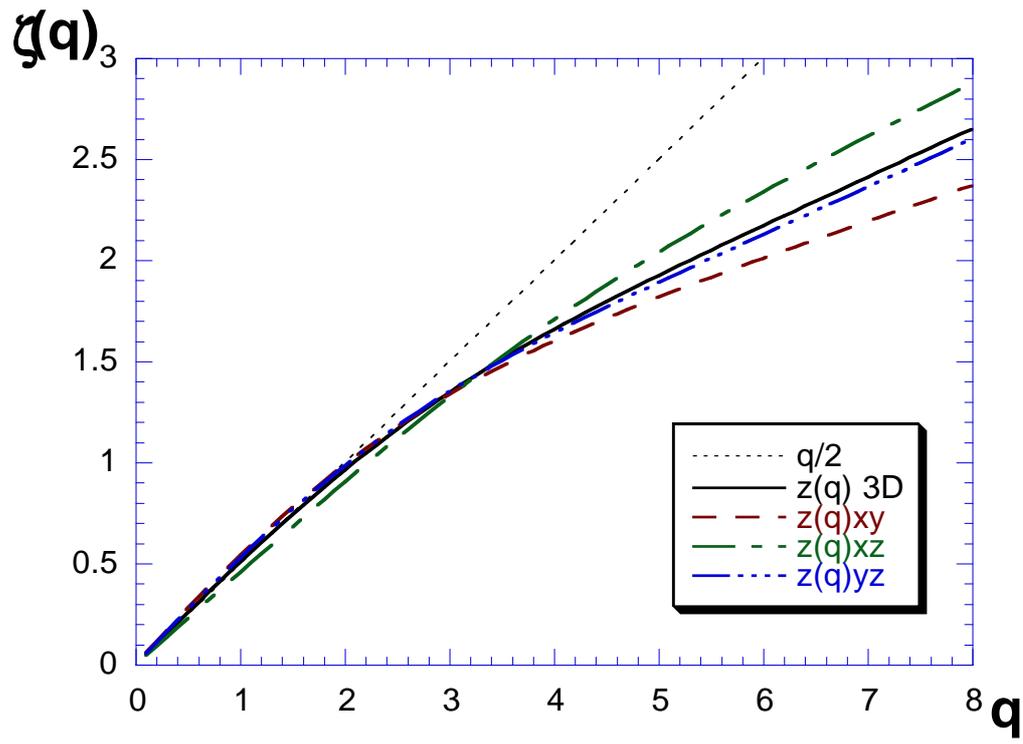

Fig. 8

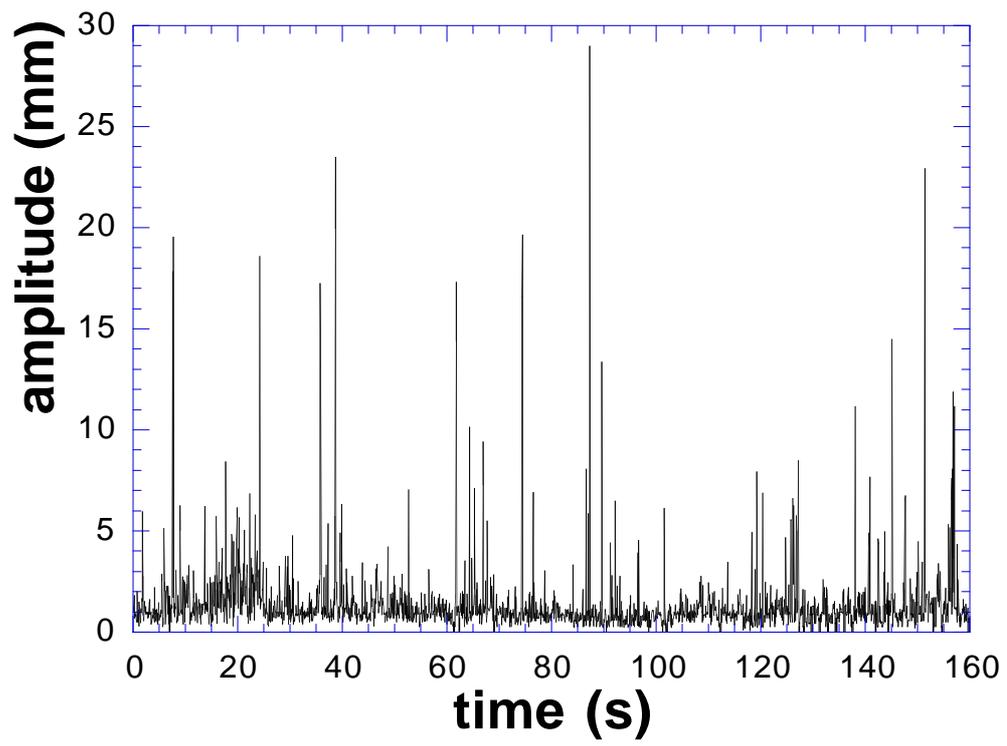

Fig. 9

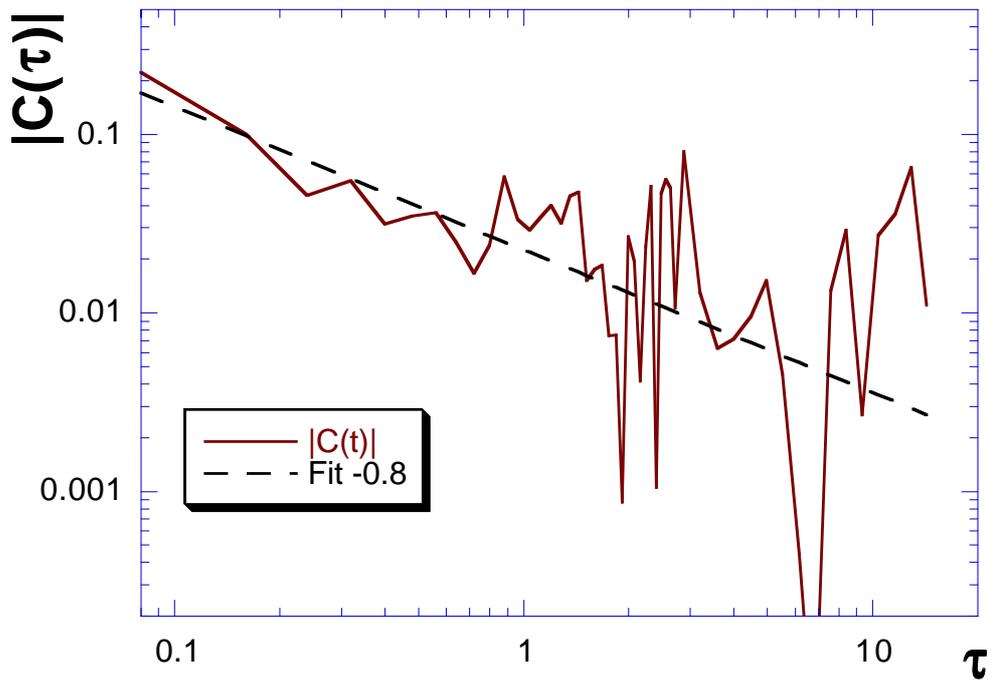

Fig. 10

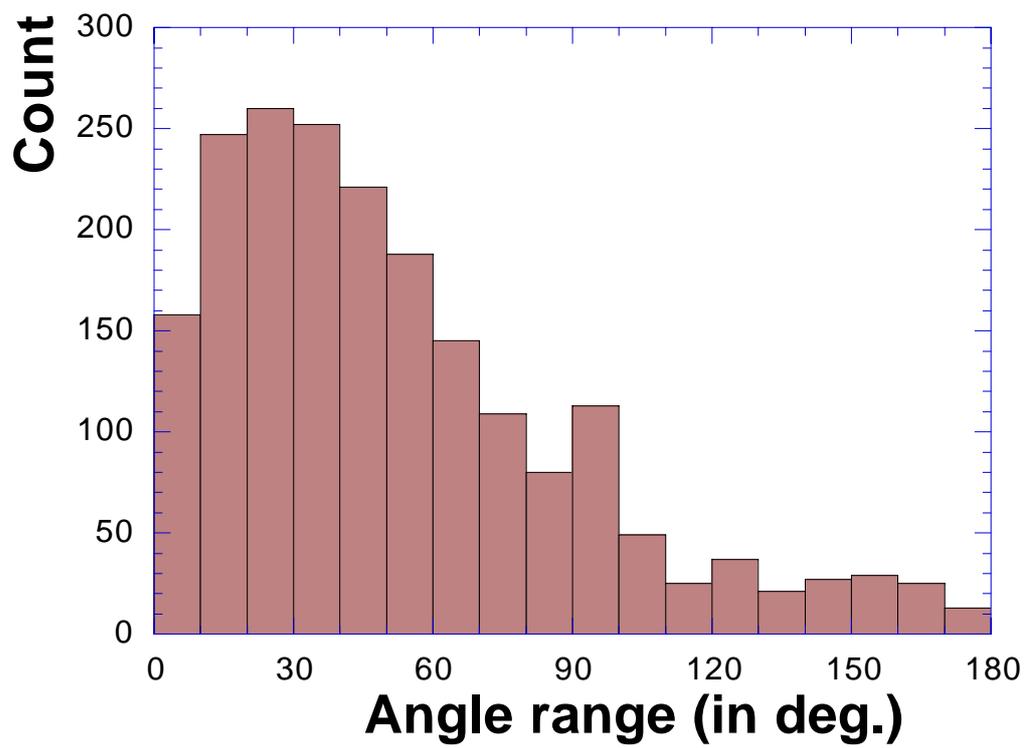

Fig. 11